%% file: main.tex
\documentclass[sigconf]{acmart}
\usepackage{xspace}
\usepackage{caption}
\usepackage{subcaption}
\usepackage{multirow}
\usepackage[normalem]{ulem}
\usepackage{subcaption}
\usepackage{xcolor}
\usepackage[linesnumbered,ruled]{algorithm2e}
\usepackage{cleveref}
\usepackage{amsmath}
\usepackage{booktabs}
\usepackage{varwidth}
\usepackage{paralist}
\usepackage{wrapfig}
\usepackage{bm}
\usepackage{multirow}
\usepackage{algorithmic}
\usepackage[normalem]{ulem}
\useunder{\uline}{\ul}{}
\usepackage{tikz}

\newcommand{\our}{\mbox{\textsc{Recformer}}\xspace}

\usepackage{enumitem}

\AtBeginDocument{%
  \providecommand\BibTeX{{%
    \normalfont B\kern-0.5em{\scshape i\kern-0.25em b}\kern-0.8em\TeX}}}

\copyrightyear{2023}
\acmYear{2023}
\setcopyright{rightsretained}
\acmConference[KDD '23]{Proceedings of the 29th ACM SIGKDD Conference on Knowledge Discovery and Data Mining}{August 6--10, 2023}{Long Beach, CA, USA}
\acmBooktitle{Proceedings of the 29th ACM SIGKDD Conference on Knowledge Discovery and Data Mining (KDD '23), August 6--10, 2023, Long Beach, CA, USA}
\acmDOI{10.1145/3580305.3599519}
\acmISBN{979-8-4007-0103-0/23/08}

\begin{document}

\title{Text Is All You Need: Learning Language Representations for Sequential Recommendation}


\author{Jiacheng Li}
\affiliation{\country{University of California, San Diego}}
\email{j9li@eng.ucsd.edu}

\author{Ming Wang}
\affiliation{\country{Amazon, United States}}
\email{mingww@amazon.com}

\author{Jin Li}
\affiliation{\country{Amazon, United States}}
\email{jincli@amazon.com}

\author{Jinmiao Fu}
\affiliation{\country{Amazon, United States}}
\email{jinnmiaof@amazon.com}

\author{Xin Shen}
\affiliation{\country{Amazon, United States}}
\email{xinshen@amazon.com}

\author{Jingbo Shang}
\affiliation{\country{University of California, San Diego}}
\email{jshang@eng.ucsd.edu}

\author{Julian McAuley}
\affiliation{\country{University of California, San Diego}}
\email{jmcauley@eng.ucsd.edu}

\renewcommand{\shortauthors}{Jiacheng and Ming, et al.}

\begin{abstract}
\input{0_abstract.tex}
\end{abstract}

\begin{CCSXML}
<ccs2012>
   <concept>
       <concept_id>10002951.10003317.10003347.10003350</concept_id>
       <concept_desc>Information systems~Recommender systems</concept_desc>
       <concept_significance>500</concept_significance>
       </concept>
 </ccs2012>
\end{CCSXML}

\ccsdesc[500]{Information systems~Recommender systems}
\keywords{sequential recommendation, language models}


\maketitle

\input{1_intro.tex}

\input{3_method.tex}

\input{4_experiments.tex}

\input{2_related_work.tex}
\input{5_conclusion.tex}

% \begin{acks}
% \end{acks}


\bibliographystyle{ACM-Reference-Format}
\balance
\bibliography{references}


\end{document}

%% file: 0_abstract.tex
Sequential recommendation aims to model dynamic user behavior from historical interactions. Existing methods rely on either explicit item IDs or 
%pretrained language representations as textual features for sequence modeling to better understand user preferences. 
%item contexts
general textual features
%\jingbo{what are the contexts? I looked at the latex comments and it seems like some textual features? If existing methods are using textual features too, I don't see an obvious advantage of the proposed method on the cold-start or transfer knowledge setting? }
for sequence modeling to understand user preferences.
While promising, these approaches still struggle to model cold-start items or transfer knowledge to new datasets. In this paper, we propose to model user preferences and item features as language representations that can be generalized to new items and datasets. To this end, we present a novel framework, named~\our, which effectively learns language representations for sequential recommendation. 
%Specifically, instead of IDs, we propose to formulate an item as a sequence of key-value attribute pairs and then obtain a ``sentence'' to represent an item. 
Specifically, we propose to formulate an item as a ``sentence'' (word sequence) by flattening item key-value attributes described by text 
%and an item sequence for a user becomes a ``sentence'' sequence. 
so that an item sequence for a user becomes a sequence of sentences.
For recommendation, \our is trained to understand the ``sentence'' sequence and retrieve the next ``sentence''.  
%\jingbo{The analogy is reasonable but in NLP, people believe that the next sentence prediction (NSP) in BERT or similar settings is not a very useful task. Not sure if it will give the reviewers an impression that this design is not very effective?}
To encode item sequences, we design a bi-directional Transformer similar to the model Longformer but with different embedding layers for sequential recommendation. 
For effective representation learning, we propose novel pretraining and finetuning methods which combine language understanding and recommendation tasks. 
%\jingbo{I didn't get why this proposed method can better address cold-start or domain transfer? The current design looks fairly conventional following the language model's usage in NLP. Is there anything significantly different? }
Therefore, \our can effectively recommend the next item based on language representations. 
%Extensive experiments conducted on six datasets demonstrate the effectiveness of \our. Especially, \our shows a strong ability of knowledge transferring in low-resource and cold-start settings.
Extensive experiments conducted on six datasets demonstrate the effectiveness of \our for sequential recommendation, especially in low-resource and cold-start settings.

%% file: 1_intro.tex
\section{Introduction}
\label{sec:intro}
\begin{figure}[t]
    \centering
    \includegraphics[width=\linewidth]{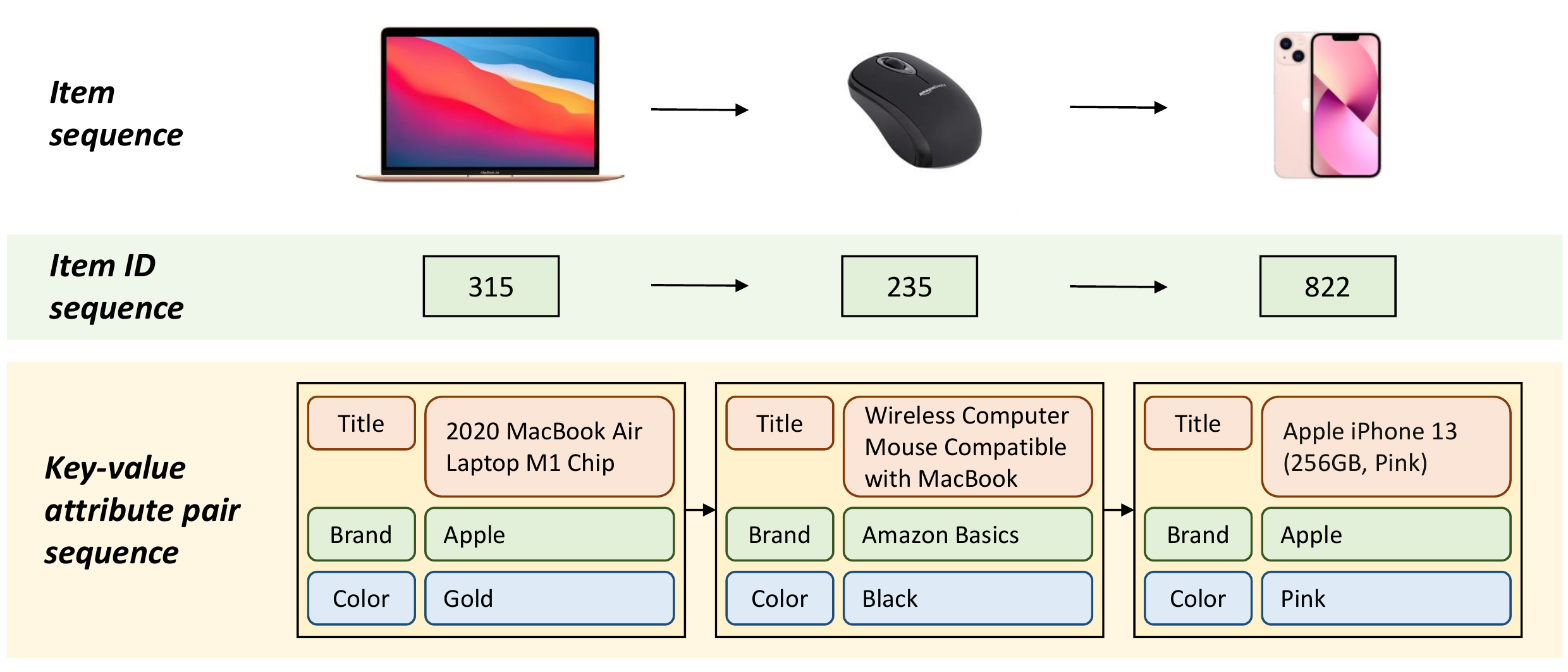}
    \caption{Input data comparison between item ID sequences for traditional sequential recommendation and key-value attribute pair sequences used in \our.}
    \label{fig:key_value_pair}
\end{figure}

Sequential recommender systems model historical user interactions as temporally-ordered sequences to recommend potential items that users are interested in. Sequential recommenders~\cite{Hidasi2015SessionbasedRW, Kang2018SelfAttentiveSR, Sun2019BERT4RecSR, Rendle2010FactorizingPM} can capture both short-term and long-term preferences of users and hence %achieve superior performance 
are widely used in different recommendation scenarios.

Various methods have been proposed to improve the performance of sequential recommendation, including Markov Chains~\cite{Rendle2010FactorizingPM, He2016FusingSM}, RNN/CNN models~\cite{Hidasi2015SessionbasedRW, Li2017NeuralAS, Tang2018PersonalizedTS, Yuan2018ASC} and self-attentive models~\cite{Sun2019BERT4RecSR, Kang2018SelfAttentiveSR, Li2020TimeIA}. Traditional sequential recommendation models convert items into IDs and create item embedding tables for encoding. Item embeddings are learned from sequences of user interactions. To enrich item features, some approaches~\cite{Zhang2019FeaturelevelDS, Zhou2020S3RecSL, Li2022CoarsetoFineSS, Chen2022IntentCL} incorporate item contexts such as item textual information or categorical features into ID embeddings. 
%\jingbo{We have just talked about some feature-based methods, it is a bit weird to go back to the ID-based methods again?}
While ID-based methods are promising, they struggle to understand cold-start items or conduct cross-domain recommendations where models are trained and then applied to different recommendation scenarios. Item-specific IDs prevent models from learning transferable knowledge from training data for cold-start items and new datasets. 
%As a result, item IDs limit the reuse of sequential recommenders across domains and we have to re-train a sequential recommender from scratch when adapting to new domains. 
%\jingbo{The limitation of ID-based methods has been well explained. How about the feature-beased ones?}
As a result, item IDs limit the performance of sequential recommenders on cold-start items and we have to re-train a sequential recommender for continually added new items. 
%\jingbo{It is not very clear to me if we should call it transferrable or something else like robust/generalizable/... I briefly checked the experiments and I don't think we have any domain transfer experiments? Mostly we are claiming something like this method works for all these datasets/domains?}
Therefore, transferable recommenders can benefit both cold-start items and new-domain datasets.

To develop transferable recommender systems, previous studies usually assume shared information such as overlapping users/items~\cite{Hu2018CoNetCC, Singh2008RelationalLV, Zhu2019DTCDRAF} and common features~\cite{Tang2012CrossdomainCR} is available and then reduce the gap between source and target domains by learning either semantic mappings~\cite{Zhu2019DTCDRAF} or transferable components~\cite{Li2021RecGURUAL}. Such assumptions are rarely true in real applications because items in different domains (e.g.,~Laptops and T-shirts) usually contain different features for  recommendation. Therefore, to have effective cross-domain transfer, recent works~\cite{Ding2021ZeroShotRS, Hou2022TowardsUS} 
%resort to 
leverage
the generality of natural language texts (e.g.,~titles, descriptions of items) for common knowledge in different domains. The basic idea is to employ pre-trained language models such as BERT~\cite{Devlin2019BERTPO} to obtain text representations and then learn the transformation from text representations to item representations. The knowledge of the transformation can be transferred across different domains and shows promising performance. However, such frameworks of learning transformation from language to items have several limitations:
%\jingbo{Now I see the limitation of the current use of language models in sequential recommendation. We should really promote this part and reduce the content for the ID-based methods. ID-based methods are too traditional. These LM-based methods are the true competitors.}
(1) Pre-trained language models are usually trained on a general language corpus (e.g.,~Wikipedia) serving natural language tasks that have a 
%totally 
different language domain from item texts (e.g.,~concatenation of item attributes), hence text representations from pre-trained language models for items are usually sub-optimal.
(2) Text representations from pre-trained language models 
%equally treat different item attributes 
are not able to learn the importance of different item attributes
and only provide coarse-grained (sentence-level) textual features but cannot learn fine-grained (word-level) user preferences for recommendations (e.g.,~find the same color in recent interactions for clothing recommendations). 
(3) Due to the independent training of pre-trained language models (by language understanding tasks, e.g.,~Masked Language Modeling) and transformation models (by recommendation tasks, e.g.,~next item prediction),  the potential ability of models to understand language for recommendations has not been fully developed (by joint training).

With the above limitations in mind, we aim to unify the frameworks of natural language understanding and recommendations in an ID-free sequential recommendation paradigm. 
The pre-trained language models~\cite{Devlin2019BERTPO, Radford2018ImprovingLU, Lewis2019BARTDS, Raffel2019ExploringTL} benefit various downstream natural language processing tasks due to their transferable knowledge from pre-training. 
%\jingbo{this previous sentence seems contradictory with the limitation (1) in the previous paragraph.}
The basic idea of this paper is to use the generality of language models through joint training of language understanding and sequential recommendations. To this end, there are three major challenges to be solved. 
%\jingbo{are these challenges related to the limitations in the previous paragraph? }
First, previous text-based methods~\cite{Hou2022TowardsUS, Ding2021ZeroShotRS} usually have their specific item texts (e.g.,~item descriptions, concatenation of item attributes). Instead of specific data types, we need to find a universal input data format of items for language models that is flexible enough to different kinds of textual item information. 
Second, it is not clear how to model languages and sequential transitions of items in one framework. Existing language models are not able to incorporate sequential patterns of items and cannot 
%figure out 
learn the alignment between items and item texts. 
Third, a training and inference framework is necessary to bridge the gap between natural languages and recommendations like how to efficiently rank items based on language models without trained item embeddings.

To address the above problems, we propose \our,  a framework that can learn language representations for sequential recommendation. Overall, our approach takes a text sequence of historical items as input and predicts the next item based on language understanding. Specifically, as shown in~\Cref{fig:key_value_pair}, we first formulate an item as key-value attribute pairs which can include any textual information such as the title, color, brand of an item. 
Different items can include different attributes as item texts. Then, to encode a sequence of key-value attribute pairs, we propose a novel bi-directional Transformer~\cite{Vaswani2017AttentionIA} based on Longformer 
%\jingbo{what is the reason of using Longformer? There are more Transformer variants that are efficient and effective on long texts?}
structure~\cite{Beltagy2020LongformerTL} but with different embeddings for item texts to learn item sequential patterns. 
Finally, to effectively learn language representations for recommendation, we design the learning framework for the model including pre-training, finetuning and inference processes. Based on the above methods, \our can effectively recommend the next items based on item text representations. Furthermore, the knowledge learned from training can be transferred to cold-start items or a new recommendation scenario.

To evaluate \our, we conduct extensive experiments on real-world datasets from different domains. Experimental results show that our method can achieve $15.83\%$ and $39.78\%$ (NDCG@10) performance improvements under fully-supervised and zero-shot sequential recommendation settings respectively.\footnote{Code will be released upon acceptance.} Our contributions in this paper can be summarized as follows:
\begin{itemize}%[leftmargin=*,nosep]
    \item We formulate items as key-value attribute pairs for the ID-free sequential recommendation and propose a novel bi-directional Transformer structure to encode sequences of key-value pairs. 
    \item We design the learning framework that helps the model learn users' preferences and then recommend items based on language representations and transfer knowledge into different recommendation domains and cold-start items.
    \item Extensive experiments are conducted to show the effectiveness of our method. Results show that \our outperforms baselines for sequential recommendation and largely improves knowledge transfer as shown by zero-shot and cold-start settings.
\end{itemize}

%% file: 3_method.tex
\section{Methodology}
\begin{figure}[t]
    \centering
    \includegraphics[width=\linewidth]{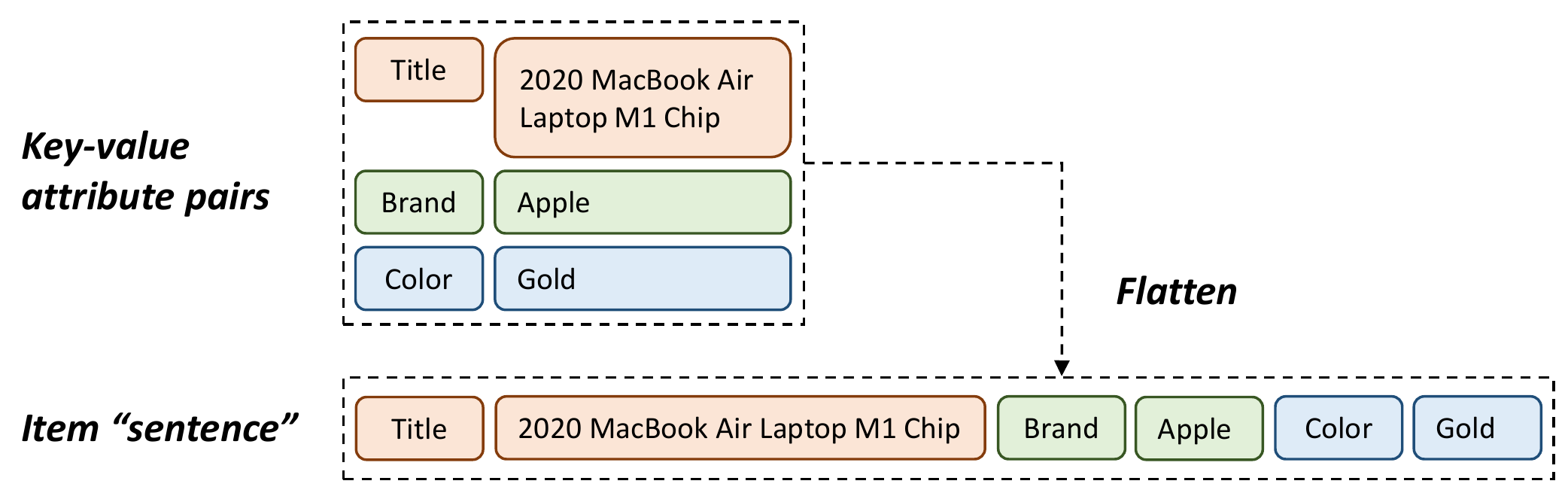}
    \caption{Model input construction. Flatten key-value attribute pairs into an item ``sentence''.}
    \label{fig:inputs}
\end{figure}

\begin{figure*}[t]
    \centering
    \includegraphics[width=0.9\linewidth]{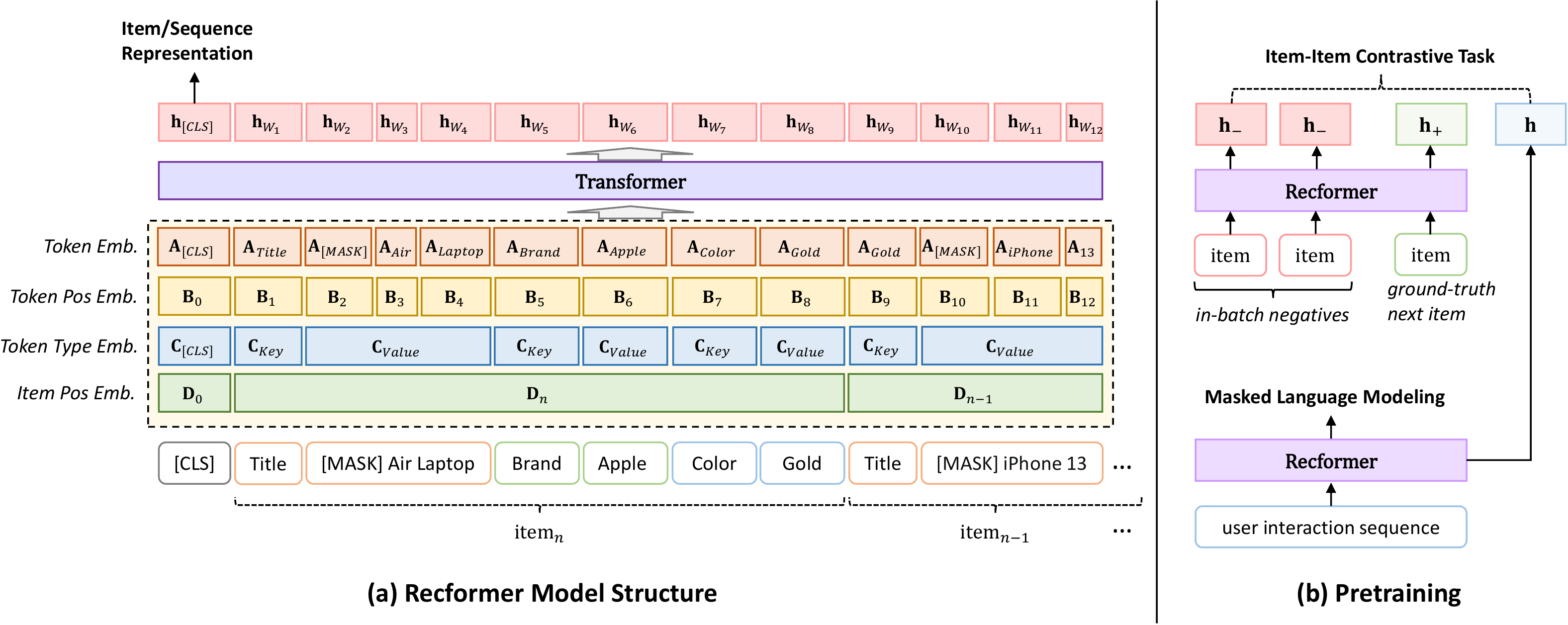}
    \vspace{-3mm}
    \caption{The overall framework of \our.}
    \vspace{-3mm}
    \label{fig:overall}
\end{figure*}

In this section, we present \our which can learn language representations for sequential recommendation and effectively transfer and generalize to new recommendation scenarios.

\subsection{Problem Setup and Formulation}
\label{sec:setup}
In the setting of sequential recommendation, we are given an item set $\mathcal{I}$ and a user's interaction sequence $s=\{i_1, i_2,\dots,i_n\}$ in temporal order where $n$ is the length of $s$ and $i\in\mathcal{I}$. Based on $s$, we seek to predict the next item. In previous sequential recommendation methods, each interacted item $i$ is associated with a unique item ID. In this paper, each item $i$ has a corresponding attribute dictionary $D_i$ containing key-value attribute pairs $\{(k_1, v_1), (k_2, v_2), \dots, (k_m, v_m)\}$ where $k$ denotes an attribute name (e.g.,~\textit{Color}) and $v$ is the corresponding value (e.g.,~\textit{Black}). $k$ and $v$ are both described by natural languages and contain words $(k, v) = \{w^k_1,\dots,w^k_c, w^v_1,\dots,w^v_c\}$, where $w^k$ and $w^v$ are words of $k$ and $v$ from a shared vocabulary in the language model and $c$ denotes the truncated length of text.
An attribute dictionary $D_i$ can include all kinds of item textual information such as titles, descriptions, colors, etc. As shown in~\Cref{fig:inputs}, to feed the attribute dictionary $D_i$ into a language model, we flatten key-value attribute pairs into $T_i = \{k1, v1, k2, v2, \dots, k_m, v_m\}$ to obtain an item ``sentence'' as the input data. Unlike previous sequential recommenders~\cite{Hou2022TowardsUS, Zhang2019FeaturelevelDS} using both text and item IDs, in this study, we use \textit{only text} for the sequential recommendation.

\subsection{Recformer}
\Cref{fig:overall} (a) shows the architecture of \our. The model has a similar structure as Longformer~\cite{Beltagy2020LongformerTL} which adopts a multi-layer bidirectional Transformer~\cite{Vaswani2017AttentionIA} with an attention mechanism that scales linearly with sequence length. We consider only computing efficiency for using Longformer but our method is open to other bidirectional Transformer structures such as BERT~\cite{Devlin2019BERTPO} and BigBird~\cite{Zaheer2020BigBT}. 

\subsubsection{Model Inputs}
As introduced in~\Cref{sec:setup}, for each item $i$ and corresponding attribute dictionary $D_i$, we flatten the dictionary into an item ``sentence'' $T_i = \{k1, v1, k2, v2, \dots, k_m, v_m\}$ where $k$ and $v$ are described by words, formally $(k, v) = \{w^k_1,\dots,w^k_c, w^v_1,\dots,w^v_c\}$. To encode a user's interaction sequence $s=\{i_1, i_2,\dots,i_n\}$, we first reverse items in a sequence to $\{i_n, i_{n-1},\dots,i_1\}$ because intuitively recent items (i.e.,~$i_n, i_{n-1}, \dots$) are important for the next item prediction and reversed sequences can make sure recent items are included in the input data. Then, we use the item ``sentences'' to replace items and add a special token \texttt{[CLS]} at the beginning of sequences. Hence, model inputs are denoted as:
\begin{equation}
\label{eq:X}
    X = \{\text{\texttt{[CLS]}}, T_n, T_{n-1}, \dots, T_1\}
\end{equation}
where $X$ is a sequence of words containing all items and corresponding attributes the user interacted with in the historical interactions. 

\subsubsection{Embedding Layer}
The target of \our is to understand the model input $X$ from both language understanding and sequential patterns in recommendations. The key idea in our work is to combine the embedding layers from language models~\cite{Devlin2019BERTPO, Liu2019RoBERTaAR} and self-attentive sequential recommenders~\cite{Kang2018SelfAttentiveSR, Sun2019BERT4RecSR}. Hence, \our contains four embeddings as follows:
\begin{itemize}
    \item \textbf{Token embedding} represents the corresponding tokens. We denote the word token embedding by $\mathbf{A}\in\mathbb{R}^{V_w\times d}$, where $V_w$ is the number of words in our vocabulary and $d$ is the embedding dimension. \our does not have item embeddings as previous sequential recommenders and hence \our understands items in interaction sequences mainly based on these token embeddings. The size of token embeddings is a constant for different recommendation scenarios; hence, our model size is irrelevant to the number of items.
    \item \textbf{Token position embedding} represents the position of tokens in a sequence. A word appearing at the $i$-th position in the sequence $X$ is represented as $\mathbf{B}_i \in \mathbb{R}^d$. Similar to language models, token position embedding is designed to help Transformer understand the sequential patterns of words.
    \item \textbf{Token type embedding} represents where a token comes from. Specifically, the token type embedding totally contains three vectors $\mathbf{C}_{\text{\texttt{[CLS]}}}, \mathbf{C}_{\mathit{Key}}, \mathbf{C}_{\mathit{Value}} \in \mathbb{R}^d$ to represent if a token comes from \texttt{[CLS]}, attribute keys, or attribute values respectively. 
    %Tokens from different components 
    Different types of tokens usually have different importance for the next item prediction. For example, because most items usually have the same attribute keys in a recommendation dataset, models with token type embedding will recognize repeated words from the same attribute keys.
    \item \textbf{Item position embedding} represents the position of items in a sequence. A word from attributes of the $k$-th item in the sequence $X$ is represented as $\mathbf{D}_k \in \mathbb{R}^d$ and $\mathbf{D}\in\mathbb{R}^{n\times d}$ where $n$ is the maximum length of a user's interaction sequence $s$. Same as previous self-attentive sequential recommenders~\cite{Kang2018SelfAttentiveSR, Sun2019BERT4RecSR}, the item position embedding is a key component for item sequential pattern learning. In \our, the item position embedding can also help the model learn the alignment between word tokens and items.
\end{itemize}

Therefore, given a word $w$ from the input sequence $X$, the input embedding is calculated as the summation of four different embeddings followed by  layer normalization~\cite{Ba2016LayerN}:
\begin{equation}
    \mathbf{E}_w = \mathrm{LayerNorm}(\mathbf{A}_w + \mathbf{B}_w + \mathbf{C}_w + \mathbf{D}_w)
\end{equation}
where $\mathbf{E}_w \in \mathbb{R}^d$.

The embedding of model inputs $X$ is a sequence of $\mathbf{E}_w$,
\begin{equation}
    \mathbf{E}_X = [\mathbf{E}_{\text{\texttt{[CLS]}}}, \mathbf{E}_{w_1}, \dots, \mathbf{E}_{w_l}]
\end{equation}
where $\mathbf{E}_X \in \mathbb{R}^{(l+1)\times d}$ and $l$ is the maximum length of tokens in a user's interaction sequence.

\subsubsection{Item or Sequence Representations}
\label{sec:repr}
To encode $\mathbf{E}_X$, we employ the bidirectional Transformer structure Longformer~\cite{Beltagy2020LongformerTL} as our encoder. Because $X$ is usually a long sequence, the local windowed attention in Longformer can help us efficiently encode $\mathbf{E}_X$. As the standard settings in Longformer for document understanding, the special token \texttt{[CLS]} has global attention but other tokens use the local windowed attention. Hence, \our computes $d$-dimensional word representations as follows:
\begin{equation}
    [\mathbf{h}_{\text{\texttt{[CLS]}}}, \mathbf{h}_{w_1}, \dots, \mathbf{h}_{w_l}] = \mathrm{Longformer}([\mathbf{E}_{\text{\texttt{[CLS]}}}, \mathbf{E}_{w_1}, \dots, \mathbf{E}_{w_l}])
\end{equation}
where $\mathbf{h}_w \in \mathbb{R}^d$. Similar to the language models used for sentence representations, the representation of the first token $\mathbf{h}_{\text{\texttt{[CLS]}}}$ is used as the sequence representation.

In \our, we do not maintain an embedding table for items. Instead, we view the item as a special case of the interaction sequence with only one item. For each item $i$, we construct its item ``sentence'' $T_i$ and use $X = \{\text{\texttt{[CLS]}}, T_i\}$ as the model input to get the sequence representation $\mathbf{h}_{\text{\texttt{[CLS]}}}$ as the item representation $\mathbf{h}_i$.

\subsubsection{Prediction}
We predict the next item based on the cosine similarity between a user's interaction sequence $s$ and item $i$. Formally, after obtaining the sequence representation $\mathbf{h}_s$ and the item representation $\mathbf{h}_i$ as introduced in~\Cref{sec:repr}, we calculate the scores between $s$ and $i$ as follows:
\begin{equation}
\label{eq:sim}
    r_{i, s} = \frac{\mathbf{h}_i^{\top} \mathbf{h}_s}{\Vert \mathbf{h}_i \Vert \cdot \Vert \mathbf{h}_s \Vert}
\end{equation}
where $r_{i, s} \in \mathbb{R}$ is the relevance of item $i$ being the next item given $s$. To predict the next item, we calculate $r_{i, s}$ for all items~\footnote{For efficient calculation, we encode all items in advance for score calculation.} in the item set $\mathcal{I}$ and select item with the highest $r_{i, s}$ as the next item:
\begin{equation}
    \hat{i_s} = \mathrm{argmax}_{i\in \mathcal{I}}(r_{i, s})
\end{equation}
where $\hat{i_s}$ is the predicted item given user interaction sequence $s$.

\subsection{Learning Framework}
To have an effective and efficient language model for the sequential recommendation, we propose our learning framework for \our including pre-training and two-stage finetuning.

\subsubsection{Pre-training}
The target of pre-training is to obtain a high-quality parameter initialization for downstream tasks. Different from previous sequential recommendation pre-training methods which consider only recommendations, we need to consider both language understanding and recommendations. Hence, to pre-train \our, we adopt two tasks: (1) Masked Language Modeling (MLM) and (2) an item-item contrastive task.

Masked Language Modeling (MLM)~\cite{Devlin2019BERTPO} is an effective pre-training method for language understanding and has been widely used for various NLP pre-training tasks such as sentence understanding~\cite{gao2021simcse}, phrase understanding~\cite{Li2022UCTopicUC}. Adding MLM as an auxiliary task will prevent language models from forgetting the word semantics when models are jointly trained with other specific tasks. For recommendation tasks, MLM can also eliminate the language domain gap between a general language corpus and item texts. In particular, following BERT~\cite{Devlin2019BERTPO}, the training data generator chooses $15\%$ of the token positions at random for prediction. If the token is selected, we replace the token with (1) the \texttt{[MASK]} with probability $80\%$; (2) a random token with probability $10\%$; (3) the unchanged token with probability $10\%$. The MLM loss is calculated as:
\begin{align}
    \mathbf{m} &= \mathrm{LayerNorm}(\mathrm{GELU}(\mathbf{W}_h \mathbf{h}_w + \mathbf{b}_h)) \\
    p &= \mathrm{Softmax}(\mathbf{W}_0 \mathbf{m} + \mathbf{b}_0 )\\
    \mathcal{L}_{\mathrm{MLM}} &=  - \sum_{i=0}^{|\mathcal{V}|} y_i \log(p_i)
\end{align}
where $\mathbf{W}_h \in \mathbb{R}^{d \times d}$, $\mathbf{b}_h \in \mathbb{R}^{d}$, $\mathbf{W}_0 \in \mathbb{R}^{|\mathcal{V}| \times d}$, $\mathbf{b}_0 \in \mathbb{R}^{|\mathcal{V}|}$, $\mathrm{GELU}$ is the GELU activation function~\cite{Hendrycks2016GaussianEL} and $\mathcal{V}$ is the vocabulary used in the language model.

Another pre-training task for \our is the item-item contrastive (IIC) task which is widely used in the next item prediction for recommendations. We use the ground-truth next items as positive instances following previous works~\cite{Hou2022TowardsUS, Kang2018SelfAttentiveSR, Sun2019BERT4RecSR}. However, for negative instances, we adopt in-batch next items as negative instances instead of negative sampling~\cite{Kang2018SelfAttentiveSR} or fully softmax~\cite{Sun2019BERT4RecSR, Hou2022TowardsUS}. Previous recommenders maintain an item embedding table, hence they can easily retrieve item embeddings for training and update embeddings. In our case, item embeddings are from \our, so it is infeasible to re-encode items (from sampling or full set) per batch for training. In-batch negative instances~\cite{Chen2017OnSS} are using ground truth items of other instance sequences in the same batch as  negative items. Although it is possible to provide false negatives, false negatives are less likely in the pre-training dataset with a large size. Furthermore, the target of pre-training is to provide high-quality initialized parameters and we have the finetuning with accurate supervision for downstream tasks. Therefore, we claim that in-batch negatives will not hurt the recommendation performance but have much higher training efficiency than accurate supervision. Formally, the item-item contrastive loss is calculated as:
\begin{equation}
    \mathcal{L}_{\mathrm{IIC}} = -\log \frac{e^{\mathrm{sim}(\mathbf{h}_s, \mathbf{h}_i^+)/\tau}}{\sum_{i\in \mathcal{B}}e^{\mathrm{sim}(\mathbf{h}_s, \mathbf{h}_i)/\tau}}
\end{equation}
where $\mathrm{sim}$ is the similarity introduced in~\Cref{eq:sim}; $\mathbf{h}_i^+$ is the representation of the ground truth next item; $\mathcal{B}$ is the ground truth item set in one batch and $\tau$ is a temperature parameter.

At the pre-training stage, we use a multi-task training strategy to jointly optimize \our:
\begin{equation}
    \mathcal{L}_{\mathrm{PT}} = \mathcal{L}_{\mathrm{IIC}} + \lambda \cdot \mathcal{L}_{\mathrm{MLM}}
\end{equation}
where $\lambda$ is a hyper-parameter to control the weight of MLM task loss. The pre-trained model will be fine-tuned for new scenarios.

\begin{algorithm}[t]
\caption{Two-Stage Finetuning}
\label{alg:finetune}

\textbf{Input}: $D_{\mathrm{train}}$, $D_{\mathrm{valid}}$, $\mathcal{I}$, $M$ \\
\textbf{Hyper-parameters}: $n_{\mathrm{epoch}}$ \\
\textbf{Output}: $M^\prime$, $\mathbf{I}^\prime$

\begin{algorithmic}[1]
\STATE $M \leftarrow$ initialized with pre-trained parameters \\
\STATE $p \leftarrow$ metrics are initialized with 0 \\

\emph{Stage 1} \\

\FOR{$n$ in $n_{\mathrm{epoch}}$}
    \STATE $\mathbf{I} \leftarrow \mathrm{Encode}(M, \mathcal{I})$ \\
    \STATE $M \leftarrow \mathrm{Train}(M, \mathbf{I}, D_{\mathrm{train}})$ \\
    \STATE $p^\prime \leftarrow \mathrm{Evaluate}(M, \mathbf{I}, D_{\mathrm{valid}})$ \\
    \IF{$p^\prime > p$}
    \STATE $M^\prime, \mathbf{I}^\prime \leftarrow M, \mathbf{I}$ \\
    \STATE $p \leftarrow p^\prime$
    \ENDIF
\ENDFOR 

\emph{Stage 2} \\

\STATE $M \leftarrow M^\prime$ \\

\FOR{$n$ in $n_{\mathrm{epoch}}$}
    \STATE $M \leftarrow \mathrm{Train}(M, \mathbf{I}^\prime, D_{\mathrm{train}})$ \\
    \STATE $p^\prime \leftarrow \mathrm{Evaluate}(M, \mathbf{I}^\prime, D_{\mathrm{valid}})$ \\
    \IF{$p^\prime > p$}
    \STATE $M^\prime \leftarrow M$
    \STATE $p \leftarrow p^\prime$
    \ENDIF
\ENDFOR 

\STATE \textbf{return} $M^\prime$, $\mathbf{I}^\prime$ \\

\end{algorithmic}
\end{algorithm} 

\subsubsection{Two-Stage Finetuning}
Similar to pre-training, we do not maintain an independent item embedding table. Instead, we encode items by \our. However, in-batch negatives cannot provide accurate supervision in a small dataset because it is likely to have false negatives which undermine recommendation performance. To solve this problem, we propose two-stage finetuning as shown in~\Cref{alg:finetune}. The key idea is to maintain an item feature matrix $\mathbf{I}\in \mathbb{R}^{|\mathcal{I}|\times d}$. Different from the item embedding table, $\mathbf{I}$ is not learnable and all item features are encoded from \our. As shown in~\Cref{alg:finetune}, our proposed finetuning method has two stages. In stage 1, $\mathbf{I}$ is updated (line 4) per epoch,\footnote{Updating means encoding all items with \our.} whereas, in stage 2 we freeze $\mathbf{I}$ and update only parameters in model $M$. The basic idea is that although the model is already pre-trained, item representations from the pre-trained model can still be improved by further training on downstream datasets. It is expensive to re-encode all items in every batch hence we re-encode all items in every epoch to update $\mathbf{I}$ (line 4) and use $\mathbf{I}$ as supervision for item-item contrastive learning (line 5). After obtaining the best item representations, we re-initialize the model with the corresponding parameters (line 12) and start stage 2. Since $\mathbf{I}$ keeps updating in stage 1, the supervision for finetuning is also changing. In this case, the model is hard to be optimized to have the best performance. Therefore, we freeze $\mathbf{I}$ and continue training the model until achieving the best performance on the validation dataset.

The learning task used in finetuning is item-item contrastive learning which is the same as pre-training but with fully softmax instead of in-batch negatives. The finetuning loss is calculated as:

\begin{equation}
    \mathcal{L}_{\mathrm{FT}} = -\log \frac{e^{\mathrm{sim}(\mathbf{h}_s, \mathbf{I}_i^+)/\tau}}{\sum_{i\in \mathcal{I}}e^{\mathrm{sim}(\mathbf{h}_s, \mathbf{I}_i)/\tau}}
\end{equation}
where $\mathbf{I}_i$ is the item feature of item $i$.

\subsection{Discussion}
In this section, we briefly compare \our to other sequential recommendation methods to highlight the novelty of our method.

\textbf{Traditional sequential recommenders} such as GRU4Rec~\cite{Hidasi2015SessionbasedRW}, SASRec~\cite{Kang2018SelfAttentiveSR} and BERT4Rec~\cite{Sun2019BERT4RecSR} rely on item IDs and corresponding trainable item embeddings to train a sequential model for recommendations. These item embeddings are learned from sequential patterns of user interactions. However, as mentioned in~\cite{Li2022CoarsetoFineSS}, these approaches suffer from data sparsity and can not perform well with cold-start items.

To reduce the dependence on item IDs, some \textbf{context-aware sequential recommenders} such as UniSRec~\cite{Hou2022TowardsUS}, S$^3$-Rec~\cite{Zhou2020S3RecSL}, ZESRec~\cite{Ding2021ZeroShotRS} are proposed to incorporate side information (e.g.,~categories, titles) as prior knowledge for recommendations. All of these approaches rely on a feature extractor such as BERT~\cite{Devlin2019BERTPO} to obtain item feature vectors and then fuse these vectors into item representations with an independent sequential model.

In this paper, we explore conducting sequential recommendations in a 
%brand 
new paradigm that learns language representations for the next item recommendations. Instead of trainable item embeddings or fixed item features from language models, we bridge the gap between natural language understanding and sequential recommendation to directly learn representations of items and user sequences based on words. We expect the generality of natural language can improve the transferability of recommenders in order to benefit new domain adaptation and cold-start item understanding.

%% file: 4_experiments.tex
\section{Experiments}
In this section, we empirically show the effectiveness of our proposed model \our and learning framework.

\begin{table}[t]
\small
\centering
\caption{Statistics of the datasets after preprocessing. Avg. n denotes the average length of item sequences.}
\label{tab:stat}
\begin{tabular}{lrrrrr}
\toprule
\multicolumn{1}{c}{\textbf{Datasets}} & \multicolumn{1}{c}{\textbf{\#Users}} & \multicolumn{1}{c}{\textbf{\#Items}} & \multicolumn{1}{c}{\textbf{\#Inters.}} & \multicolumn{1}{c}{\textbf{Avg. n}} & \multicolumn{1}{c}{\textbf{Density}} \\ \midrule
\textbf{Pre-training}                 & 3,613,906                            & 1,022,274                            & 33,588,165                             & 9.29                                & 9.1e-6                               \\
-Training                             & 3,501,527                            & 954,672                              & 32,291,280                             & 9.22                                & 9.0e-6                               \\
-Validation                           & 112,379                              & 67,602                               & 1,296,885                              & 11.54                               & 1.7e-4                               \\ \midrule
\textbf{Scientific}                   & 11,041                               & 5,327                                & 76,896                                 & 6.96                                & 1.3e-3                               \\
\textbf{Instruments}                  & 27,530                               & 10,611                               & 231,312                                & 8.40                                & 7.9e-4                               \\
\textbf{Arts}                         & 56,210                               & 22,855                               & 492,492                                & 8.76                                & 3.8e-4                               \\
\textbf{Office}                       & 101,501                              & 27,932                               & 798,914                                & 7.87                                & 2.8e-4                               \\
\textbf{Games}                        & 11,036                               & 15,402                               & 100,255                                & 9.08                                & 5.9e-4                               \\
\textbf{Pet}                          & 47,569                               & 37,970                               & 420,662                                & 8.84                                & 2.3e-4                               \\ \bottomrule
\end{tabular}
\end{table}

\begin{table*}[t]
\centering
\small
\caption{Performance comparison of different recommendation models. The best and the second-best 
%performances 
performance
is bold and underlined respectively. Improv.~denotes the relative improvement of \our over the best baselines.}
\vspace{-3mm}
\label{tab:perf}
\scalebox{0.9}{
\setlength{\tabcolsep}{1mm}{
\begin{tabular}{llcccccccccc}
\toprule
\multicolumn{1}{c}{\textbf{}} &    \multicolumn{1}{c}{\textbf{}}   & \multicolumn{4}{c}{\textbf{ID-Only Methods}}                                                                    & \multicolumn{2}{c}{\textbf{ID-Text Methods}} & \multicolumn{3}{c}{\textbf{Text-Only Methods}}          & \multirow{2}{*}{\textbf{Improv.}} \\ \cmidrule(lr){3-6} \cmidrule(lr){7-8} \cmidrule(lr){9-11}
\multicolumn{1}{c}{\textbf{Dataset}} & \multicolumn{1}{c}{\textbf{Metric}} & \textbf{GRU4Rec} & \textbf{SASRec} & \textbf{BERT4Rec} & \textbf{RecGURU} & \textbf{FDSA}        & \textbf{S$^3$-Rec}       & \textbf{ZESRec} & \textbf{UniSRec} & \textbf{\our} &                                      \\ \midrule
\multirow{3}{*}{Scientific}          & NDCG@10                             & 0.0826           & 0.0797          & 0.0790            & 0.0575           & 0.0716               & 0.0451                & 0.0843          & {\ul 0.0862}     & \textbf{0.1027}    & 19.14\%                              \\
                                     & Recall@10                           & 0.1055           & {\ul 0.1305}    & 0.1061            & 0.0781           & 0.0967               & 0.0804                & 0.1260          & 0.1255           & \textbf{0.1448}    & 10.96\%                              \\
                                     & MRR                                 & 0.0702           & 0.0696          & 0.0759            & 0.0566           & 0.0692               & 0.0392                & 0.0745          & {\ul 0.0786}     & \textbf{0.0951}    & 20.99\%                              \\ \midrule
\multirow{3}{*}{Instruments}         & NDCG@10                             & 0.0633           & 0.0634          & 0.0707            & 0.0468           & 0.0731               & {\ul 0.0797}          & 0.0694          & 0.0785           & \textbf{0.0830}    & 4.14\%                               \\
                                     & Recall@10                           & 0.0969           & 0.0995          & 0.0972            & 0.0617           & 0.1006               & {\ul 0.1110}          & 0.1078          & \textbf{0.1119}  & 0.1052             & -                                    \\
                                     & MRR                                 & 0.0707           & 0.0577          & 0.0677            & 0.0460           & 0.0748               & {\ul 0.0755}          & 0.0633          & 0.0740           & \textbf{0.0807}    & 6.89\%                               \\ \midrule
\multirow{3}{*}{Arts}                & NDCG@10                             & {\ul 0.1075}     & 0.0848          & 0.0942            & 0.0525           & 0.0994               & 0.1026                & 0.0970          & 0.0894           & \textbf{0.1252}    & 16.47\%                              \\
                                     & Recall@10                           & 0.1317           & 0.1342          & 0.1236            & 0.0742           & 0.1209               & {\ul 0.1399}          & 0.1349          & 0.1333           & \textbf{0.1614}    & 15.37\%                              \\
                                     & MRR                                 & 0.1041           & 0.0742          & 0.0899            & 0.0488           & 0.0941               & {\ul 0.1057}          & 0.0870          & 0.0798           & \textbf{0.1189}    & 12.49\%                              \\ \midrule
\multirow{3}{*}{Office}              & NDCG@10                             & 0.0761           & 0.0832          & {\ul 0.0972}      & 0.0500           & 0.0922               & 0.0911                & 0.0865          & 0.0919           & \textbf{0.1141}    & 17.39\%                              \\
                                     & Recall@10                           & 0.1053           & 0.1196          & 0.1205            & 0.0647           & {\ul 0.1285}         & 0.1186                & 0.1199          & 0.1262           & \textbf{0.1403}    & 9.18\%                               \\
                                     & MRR                                 & 0.0731           & 0.0751          & 0.0932            & 0.0483           & {\ul 0.0972}         & 0.0957                & 0.0797          & 0.0848           & \textbf{0.1089}    & 12.04\%                              \\ \midrule
\multirow{3}{*}{Games}               & NDCG@10                             & 0.0586           & 0.0547          & {\ul 0.0628}      & 0.0386           & 0.0600               & 0.0532                & 0.0530          & 0.0580           & \textbf{0.0684}    & 8.92\%                               \\
                                     & Recall@10                           & 0.0988           & 0.0953          & {\ul 0.1029}      & 0.0479           & 0.0931               & 0.0879                & 0.0844          & 0.0923           & \textbf{0.1039}    & 0.97\%                               \\
                                     & MRR                                 & 0.0539           & 0.0505          & {\ul 0.0585}      & 0.0396           & 0.0546               & 0.0500                & 0.0505          & 0.0552           & \textbf{0.0650}    & 11.11\%                              \\ \midrule
\multirow{3}{*}{Pet}                 & NDCG@10                             & 0.0648           & 0.0569          & 0.0602            & 0.0366           & 0.0673               & 0.0742                & {\ul 0.0754}    & 0.0702           & \textbf{0.0972}    & 28.91\%                              \\
                                     & Recall@10                           & 0.0781           & 0.0881          & 0.0765            & 0.0415           & 0.0949               & {\ul 0.1039}          & 0.1018          & 0.0933           & \textbf{0.1162}    & 11.84\%                              \\
                                     & MRR                                 & 0.0632           & 0.0507          & 0.0585            & 0.0371           & 0.0650               & {\ul 0.0710}          & 0.0706          & 0.0650           & \textbf{0.0940}    & 32.39\%                              \\ \bottomrule
                                     
\end{tabular}
}}
\end{table*}

\subsection{Experimental Setup}
\subsubsection{Datasets}
To evaluate the performance of \our, we conduct pre-training and finetuning on different categories of Amazon review datasets~\cite{Ni2019JustifyingRU}. The statistics of datasets after preprocessing are shown in~\Cref{tab:stat}. 

\textbf{For pre-training}, seven categories are selected as training data including ``\textit{Automotive}'', ``\textit{Cell Phones and Accessories}'', ``\textit{Clothing Shoes and Jewelry}'', ``\textit{Electronics}'', ``\textit{Grocery and Gourmet Food}'', ``\textit{Home and Kitchen}'', ``\textit{Movies and TV}'', and one category ``\textit{CDs and Vinyl}'' is left out as validation data. Datasets from these categories are used as source domain datasets.

\textbf{For finetuning}, we select six categories including ``\textit{Industrial and Scientific}'', ``\textit{Musical Instruments}'', ``\textit{Arts, Crafts and Sewing}'', ``\textit{Office Products}'', ``\textit{Video Games}'', ``\textit{Pet Supplies}'', as target domain datasets to evaluate \our.

For pre-training and finetuning, we use the five-core datasets provided by the data source and filter items whose \emph{title} is missing. Then we group the interactions by users and sort them by timestamp ascendingly. Following previous work~\cite{Hou2022TowardsUS}, we select item attributes \emph{title}, \emph{categories} and \emph{brand} as key-value pairs for items.

\subsubsection{Baselines}
We compare three groups of works as our baselines which include methods with only item IDs; methods using item IDs and treating item text as side information; and methods using only item texts as inputs.

(1) ID-Only methods:
\begin{itemize}
    \item \textbf{GRU4Rec}~\cite{Hidasi2015SessionbasedRW} adopts RNNs to model user action sequences for session-based recommendations. We treat each user's interaction sequence as a session.
    \item \textbf{SASRec}~\cite{Kang2018SelfAttentiveSR} uses a directional self-attentive model to capture item correlations within a sequence.
    \item \textbf{BERT4Rec}~\cite{Sun2019BERT4RecSR} employs a bi-directional self-attentive model with the cloze objective for modeling user behavior sequences.
    \item \textbf{RecGURU}~\cite{Li2021RecGURUAL} proposes to pre-train sequence representations with an autoencoder in an adversarial learning paradigm. We do not consider overlapped users for this method in our setting.
\end{itemize}
(2) ID-Text methods:
\begin{itemize}
    \item \textbf{FDSA}~\cite{Zhang2019FeaturelevelDS} uses a self-attentive model to capture item and feature transition patterns.
    \item \textbf{S}$^3$\textbf{-Rec}~\cite{Zhou2020S3RecSL} pre-trains sequential models with mutual information maximization to learn the correlations among attributes, items, subsequences, and sequences.
\end{itemize}
(3) Text-Only methods:
\begin{itemize}
    \item \textbf{ZESRec}~\cite{Ding2021ZeroShotRS} encodes item texts with a pre-trained language model as item features. We pre-train this method and finetune the model on six downstream datasets.
    \item \textbf{UniSRec}~\cite{Hou2022TowardsUS} uses textual item representations from a pre-trained language model and adapts to a new domain using an MoE-enhance adaptor. We initialize the model with the pre-trained parameters provided by the authors and finetune the model on target domains.
\end{itemize}

\subsubsection{Evaluation Settings}
To evaluate the performance of sequential recommendation, we adopt three widely used metrics NDCG@N, Recall@N and MRR, where N is set to 10. For data splitting of finetuning datasets, we apply the leave-one-out strategy~\cite{Kang2018SelfAttentiveSR} for evaluation: the most recent item in an interaction sequence is used for testing, the second most recent item for validation and the remaining data for training. We rank the ground-truth item of each sequence among all items for evaluation and report the average scores of all sequences in the test data.

\subsubsection{Implementation Details}
We build \our based on Longformer implemented by Huggingface~\footnote{\url{https://huggingface.co/docs/transformers/}}. For efficient computing, we set the size of the local attention windows in Longformer to $64$. The maximum number of tokens is $32$ for each attribute and 1,024 for each interaction sequence (i.e.,~$X$ in~\Cref{eq:X}). The maximum number of items in a user sequence is $50$ for all baselines and \our. The temperature parameter $\tau$ is $0.05$ and the weight of MLM loss $\lambda$ is $0.1$. Other than token type embedding and item position embedding in \our, other parameters are initialized with pre-trained parameters of Longformer~\footnote{\url{https://huggingface.co/allenai/longformer-base-4096}} before pre-training. The batch size is 64 for pre-training and 16 for finetuning. We optimize \our with Adam optimizer with learning rate 5e-5 and adopt early stop with the patience of 5 epochs to prevent overfitting. For baselines, we use the suggested settings introduced in~\cite{Hou2022TowardsUS}.

\subsection{Overall Performance}
We compare \our to baselines on six datasets across different recommendation domains. Results are shown in~\Cref{tab:perf}.

For baselines, ID-Text methods (i.e.,~FDSA and S$^3$-Rec) achieve better results compared to ID-Only and Text-Only methods in general. Because ID-Text methods include item IDs and content features, they can learn both content-based information and sequential patterns from finetuning. Comparing Text-Only methods and ID-Only methods, we can find that on the Scientific, Instruments, and Pet datasets, Text-Only methods perform better than ID-Only methods. A possible reason is that the item transitions in these three datasets are highly related to item texts (i.e.,~title, brand) hence text-only methods can recommend the next item based on content similarity.

Our proposed method \our, achieves the best overall performance on all datasets except the Recall@10 of Instruments. \our improves the NDCG@10 by $15.83\%$  and MRR by $15.99\%$ on average over the second best results. Different from baselines, \our learns language representations for sequential recommendation without pre-trained language models or item IDs. With two-stage finetuning, \our can be effectively adapted to downstream domains and transferred knowledge from pre-training can consistently benefit finetuning tasks. The results illustrate the effectiveness of the proposed \our.
\subsection{Low-Resource Performance}
\begin{figure}
    \centering
    \includegraphics[width=0.9\linewidth]{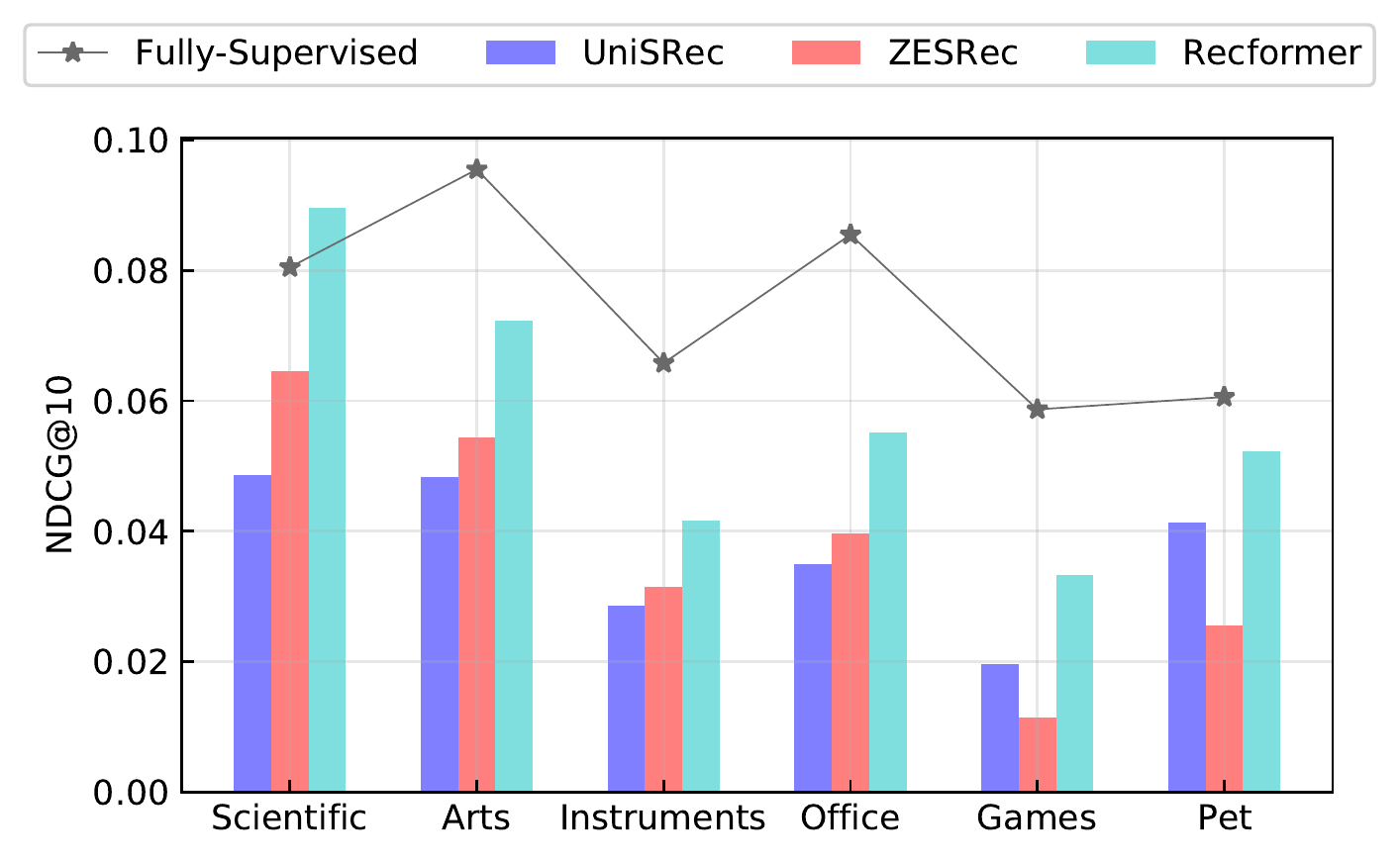}
    \vspace{-3mm}
    \caption{Performance (NDCG@10) of three Text-Only methods under the zero-shot setting. Fully-Supervised denotes the average scores of three classical ID-Only methods (i.e.,~SASRec, BERT4Rec, GRU4Rec) trained with all training data.}
    \vspace{-3mm}
    \label{fig:zero-shot}
\end{figure}

\subsubsection{Zero-Shot}
To show the effectiveness of pre-training, we evaluate the zero-shot recommendation performance of three Text-Only methods (i.e.,~UniSRec, ZESRec, \our) and compare results to the average scores of three ID-Only methods fully trained on downstream datasets. The zero-shot recommendation setting requires models to learn knowledge from pre-training datasets and directly test on downstream datasets without further training. Hence, traditional ID-based methods cannot be evaluated in this setting. We evaluate the knowledge transferability of Text-Only methods in different recommendation scenarios. All results in six downstream datasets are shown in~\Cref{fig:zero-shot}. Overall, \our improves the zero-shot recommendation performance compared to UniSRec and ZESRec on six datasets. On the Scientific dataset, \our performs better than the average performance of three ID-Only methods trained with full training sets. These results show that (1) natural language is promising 
%to work 
as a general item representation across different recommendation scenarios; (2) \our can effectively learn knowledge from pre-training and transfer learned knowledge to downstream tasks based on language understanding. 

\begin{figure}
    \centering
    \includegraphics[width=\linewidth]{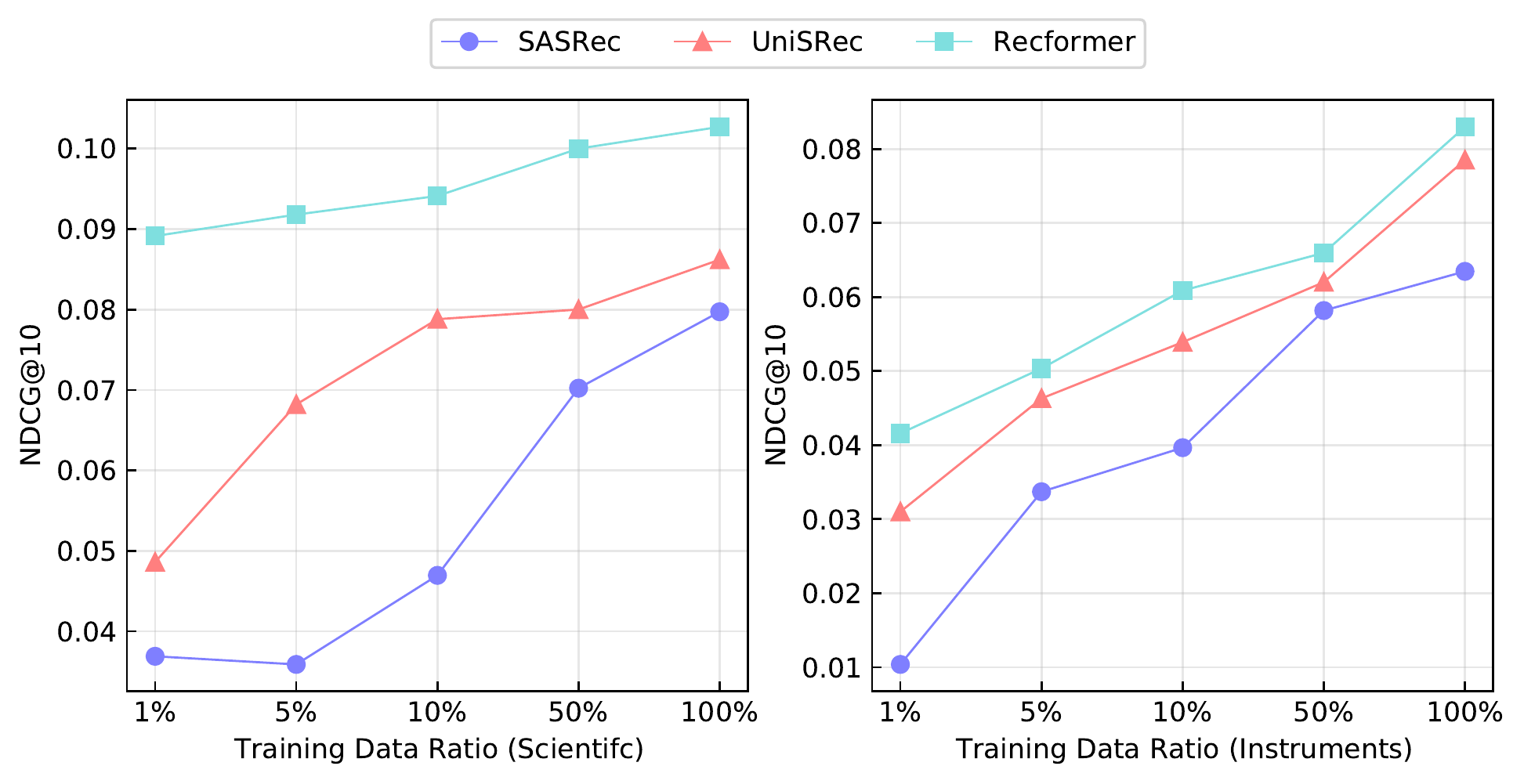}
    \vspace{-3mm}
    \caption{Performance (NDCG@10) of SASRec, UniSRec, \our over different sizes (i.e.,~$1\%$, $5\%$, $10\%$, $50\%$, $100\%$) of training data.}
    \vspace{-3mm}
    \label{fig:low_res}
    \vspace{-4mm}
\end{figure}

\subsubsection{Low-Resource}
We conduct experiments with SASRec, UniSRec and \our in low-resource settings. In this setting, we train models on downstream datasets with different ratios of training data and results are shown in~\Cref{fig:low_res}. We can see that methods with item text (i.e.,~UniSRec and \our) outperform ID-only method SASRec especially when less training data is available. This indicates UniSRec and \our can incorporate prior knowledge and do recommendations based on item texts. In low-resource settings, most items in the test set are unseen during training for SASRec. Therefore, the embeddings of unseen items are randomly initialized and cannot provide high-quality representations for recommendations. After being trained with adequate data, SASRec could rapidly improve its performance. \our achieves the best performance over different ratios of training data. On the Scientific dataset, \our outperforms other methods by a large margin with $1\%$ and $5\%$ of training data.

\subsection{Further Analysis}
\subsubsection{Performance w.r.t. Cold-Start Items}
\begin{table}[t]
\centering
\small
\caption{Performance of models compared between in-set items and cold-start items on four datasets. N@10 and R@10 stand for NDCG@10 and Recall@10 respectively.}
\vspace{-3mm}
\label{tab:cold}
\scalebox{1}{
\setlength{\tabcolsep}{1mm}{
\begin{tabular}{llcccccc}
\toprule
\multicolumn{1}{c}{\textbf{}}        & \multicolumn{1}{c}{\textbf{}}       & \multicolumn{2}{c}{\textbf{SASRec}} & \multicolumn{2}{c}{\textbf{UniSRec}} & \multicolumn{2}{c}{\textbf{\our}} \\ \cmidrule(l){3-8} 
\multicolumn{1}{c}{\textbf{Dataset}} & \multicolumn{1}{c}{\textbf{Metric}} & \textbf{In-Set}   & \textbf{Cold}   & \textbf{In-Set}    & \textbf{Cold}   & \textbf{In-Set}     & \textbf{Cold}    \\ \midrule
\multirow{2}{*}{Scientific}          & N@10                                & 0.0775            & 0.0213          & 0.0864             & 0.0441          & 0.1042              & 0.0520           \\
                                     & R@10                                & 0.1206            & 0.0384          & 0.1245             & 0.0721          & 0.1417              & 0.0897           \\ \midrule
\multirow{2}{*}{Instruments}         & N@10                                & 0.0669            & 0.0142          & 0.0715             & 0.0208          & 0.0916              & 0.0315           \\
                                     & R@10                                & 0.1063            & 0.0309          & 0.1094             & 0.0319          & 0.1130              & 0.0468           \\ \midrule
\multirow{2}{*}{Arts}                & N@10                                & 0.1039            & 0.0071          & 0.1174             & 0.0395          & 0.1568              & 0.0406           \\
                                     & R@10                                & 0.1645            & 0.0129          & 0.1736             & 0.0666          & 0.1866              & 0.0689           \\ \midrule
\multirow{2}{*}{Pet}                 & N@10                                & 0.0597            & 0.0013          & 0.0771             & 0.0101          & 0.0994              & 0.0225           \\
                                     & R@10                                & 0.0934            & 0.0019          & 0.1115             & 0.0175          & 0.1192              & 0.0400           \\ \bottomrule
\end{tabular}}}
\vspace{-4mm}
\end{table}

\begin{table*}[t]
\small
\centering
\caption{Ablation study on two downstream datasets. The best and the second-best scores are bold and underlined respectively.}
\vspace{-2mm}
\label{tab:ablation}
\scalebox{1}{
\setlength{\tabcolsep}{1mm}{
\begin{tabular}{lcccccc}
\toprule
\multicolumn{1}{c}{\multirow{2}{*}{\textbf{Variants}}} & \multicolumn{3}{c}{\textbf{Scientific}}                 & \multicolumn{3}{c}{\textbf{Instruments}}                \\ \cmidrule(l){2-4} \cmidrule(l){5-7} 
\multicolumn{1}{c}{}                                   & \textbf{NDCG@10} & \textbf{Recall@10} & \textbf{MRR}    & \textbf{NDCG@10} & \textbf{Recall@10} & \textbf{MRR}    \\ \midrule
(0) \our                                          & \textbf{0.1027}  & \textbf{0.1448}    & \textbf{0.0951} & \textbf{0.0830}  & \textbf{0.1052}    & \textbf{0.0807} \\ \cmidrule(r){1-1}
(1) w/o two-stage finetuning                           & 0.1023           & {\ul 0.1442}       & {\ul 0.0948}    & 0.0728           & 0.1005             & 0.0685          \\
(1) + (2) freezing word emb. \& item emb.              & {\ul 0.1026}     & 0.1399             & 0.0942          & 0.0728           & {\ul 0.1015}       & 0.0682          \\
(1) + (3) trainable word emb. \& item emb.             & 0.0970           & 0.1367             & 0.0873          & {\ul 0.0802}     & {\ul 0.1015}       & 0.0759          \\
(1) + (4) trainable item emb. \& freezing word emb.    & 0.0965           & 0.1383             & 0.0856          & 0.0801           & 0.1014             & {\ul 0.0760}    \\ \cmidrule(r){1-1}
(5) w/o pre-training                                    & 0.0722           & 0.1114             & 0.0650          & 0.0598           & 0.0732             & 0.0584          \\
(6) w/o item position emb. \& token type emb.                        & 0.1018           & 0.1427             & 0.0945          & 0.0518           & 0.0670             & 0.0501          \\ \bottomrule
\end{tabular}
}}
\vspace{-3mm}
\end{table*}

%New items keep being added to the item set of online platforms, hence recommender systems that can effectively recommend new items are essential to real-world scenarios. 
In this section, we simulate this scenario by splitting a dataset into two parts, i.e.,~an in-set dataset and cold-start dataset. Specifically, for the in-set dataset, we make sure all test items appear in the training data and all other test items (never appearing in training data) will be sent to the cold-start dataset. We train models on in-set datasets and test on both in-set and cold-start datasets. In this case, models never see the cold-start items during training and item embedding tables do not contain cold-start items. We compare the ID-only method SASRec and the Text-only method UniSRec to \our. 
%For SASRec, the embeddings of cold-start items are randomly initialized. 
For ID-based SASRec, we substitute items appearing only once in the training set with a cold token and after training, we add this cold token embedding to cold-start item embeddings to provide prior knowledge~\footnote{We try to provide a reasonable method for ID-based baselines with cold-start items.}.
For UniSRec, cold-start items are represented by item texts and encoded by BERT which is identical to seen items. \our directly encode item texts to represent cold-start items.

Experimental results are shown in~\Cref{tab:cold}. We can see that Text-Only methods significantly outperform SASRec, especially on datasets with a large size (i.e.,~Arts, Pet). Because of randomly initialized cold-start item representations, the performance of SASRec is largely lower on cold-start items than in-set items. Hence, ID-only methods are not able to handle cold-start items and applying text is a promising direction. For Text-only methods, \our greatly improves performance on both in-set and cold-start datasets compared to UniSRec which indicates learning language representations is superior to obtaining text features for recommendations. 
%Note that there is a large performance gap between in-set datasets and cold-start datasets. Cold-start problems are still a challenging problem for recommendations.

\subsubsection{Ablation Study}
We analyze how our proposed components influence the final sequential recommendation performance. The results are shown in~\Cref{tab:ablation}. We introduce the variants and analyze their results respectively.

We first test the effectiveness of our proposed two-stage finetuning.
In variant (1) w/o two-stage finetuning, we do not update item feature matrix $\mathbf{I}$ and only conduct finetuning based on $\mathbf{I}$ from pre-trained parameters. We find that compared to (0) \our, (1) has similar results on Scientific but has a large margin on Instruments since the pre-trained model has better pre-trained item representations on Scientific compared to Instruments (shown in~\Cref{fig:zero-shot}). Hence, our proposed two-stage finetuning can effectively improve the sub-optimal item representations from pre-training and further improve performance on downstream datasets.

Then, we investigate the effects of freezing/trainable word embeddings and item embeddings. In our default setting (1), we freeze the item feature matrix $\mathbf{I}$ and train word embeddings of \our. In variants (2)(3)(4), we try to train the item feature matrix or freeze word embeddings. Overall, on the Scientific dataset, the model with fixed item embeddings performs better than the model with trainable item embeddings, whereas on the Instruments dataset, our model performs well when item embeddings are trainable. The divergence can be eliminated by our two-stage finetuning strategy.

Variant (5) w/o pre-training finetunes \our from scratch. We can see that (0) \our significantly outperforms Variant (5) in both datasets because without pre-training, the item feature matrix $\mathbf{I}$ is not trained and cannot provide informative supervision during finetuning even if we update $\mathbf{I}$ by two-stage finetuning. These results show the effectiveness of pre-training.

Finally, we explore the effectiveness of our proposed model structure (i.e.,~item position embeddings and token type embeddings). Variant (6) removes the two embeddings and results show that the model in (6) causes performance decay on the instruments dataset which indicates the two embeddings are necessary when the gap between pre-training and finetuning is large.

\begin{figure}
    \centering
    \includegraphics[width=\linewidth]{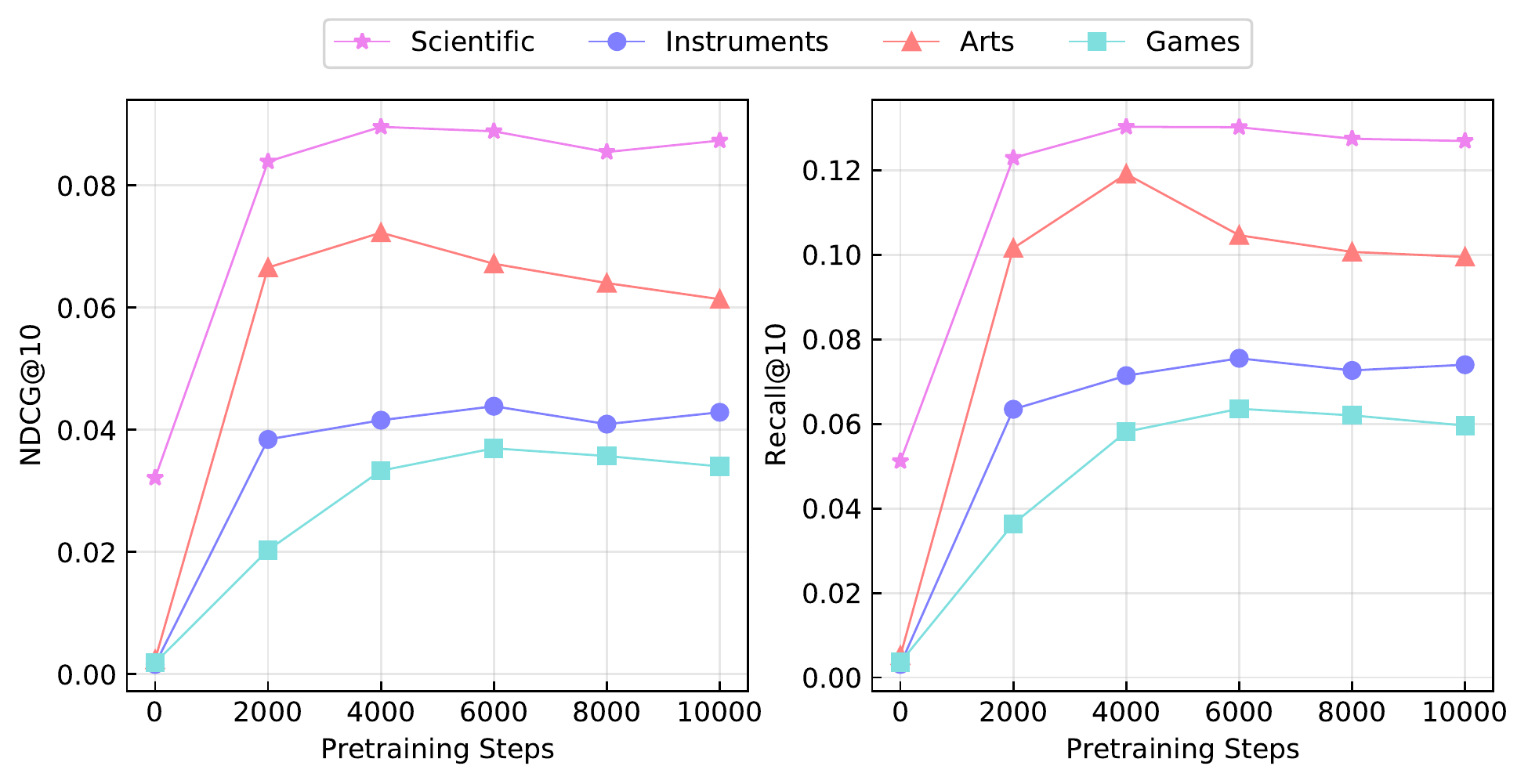}
    \vspace{-3mm}
    \caption{\our zero-shot recommendation performance (NDCG@10 and Recall@10) over different pre-training steps.}
    \label{fig:training_steps}
    \vspace{-3mm}
\end{figure}

\subsubsection{Pre-training Steps vs.~Performance}
We investigate the zero-shot sequential recommendation performance on downstream tasks over different pre-training steps and results on four datasets are shown in~\Cref{fig:training_steps}. The pre-training of natural language understanding usually requires a large number of training steps to achieve a promising result. However, we have a different situation in sequential recommendation. From~\Cref{fig:training_steps}, we can see that most datasets already achieve their best performance after around 4,000 training steps and further pre-training may hurt the knowledge transferability on downstream tasks. We think there are two possible reasons:
(1) We initialize most parameters from a Longformer model pre-trained by the MLM task. In this case, the model already has some essential knowledge of natural languages. The domain adaptation from a general language understanding to the item text understanding for recommendations should be fast.
(2) Even if we include seven categories in the training data, there is still a language domain difference between pre-training data and downstream data since different item categories have their own specific vocabularies. For instance, the category \emph{Electronics} has quite different words in item text compared to the \emph{Pets} category. 
%This finding inspires us some domain adaptation methods such as co-training of the MLM task between source domain sequences and target domain items can be explored in future studies.

%% file: 2_related_work.tex
\section{Related Work}
\subsection{Sequential Recommendation}
Sequential recommendation~\cite{Kang2018SelfAttentiveSR, Sun2019BERT4RecSR, Hidasi2015SessionbasedRW} aims to predict the next item based on historical user interactions. Proposed methods model user interactions as a sequence ordered by their timestamps. Due to the ability to capture the long-term preferences and short-term dynamics of users, sequential recommendation methods show their effectiveness for personalization and attract a lot of studies. Early works~\cite{Rendle2010FactorizingPM, He2016FusingSM} apply the Markov Chain to model item-item transition relations based on matrix factorization. For deep learning methods, Convolutional Sequence Embedding (Caser)~\cite{Tang2018PersonalizedTS} views the embedding matrix of previous items as an ``image'' and applies convolutional operations to extract transitions. GRU4Rec~\cite{Hidasi2015SessionbasedRW} introduces Gated Recurrent Units (GRU)~\cite{Chung2014EmpiricalEO} to model user sequential patterns. With the development of the Transformer~\cite{Vaswani2017AttentionIA}, recent studies~\cite{Kang2018SelfAttentiveSR, Sun2019BERT4RecSR} widely use self-attention model for sequential recommendation. Although these approaches achieve promising performance, they struggle to learn transferable knowledge or understand cold-start items due to the dependence on IDs and item embeddings which are specific to items and datasets. Recently, researchers attempt to employ textual features as transferable item representations~\cite{Ding2021ZeroShotRS, Hou2022TowardsUS}. These methods first obtain item features by encoding item texts with language models and then learn transferable item representations with an independent sequential model. Independent language understanding and sequential pattern learning still limit the capacity of the model to learn user interactions based on languages. In this paper, we explore unifying the language understanding and sequential recommendations into one Transformer framework. We aim to have a sequential recommendation method that can effectively model cold-start items and learn transferable sequential patterns for different recommendation scenarios.

\subsection{Transfer Learning for Recommendation}
Data sparsity and cold-start item understanding issues are challenging in recommender systems and recent studies~\cite{Xie2021ContrastiveCR, Zhu2019DTCDRAF, Zhu2021CrossDomainRC} explore transferring knowledge across different domains to improve the recommendation at the target domain. Previous methods for knowledge transfer mainly rely on shared information between the source and target domains including common users~\cite{Hu2018CoNetCC, Wu2020PTUMPU, Xiao2021UPRecUP, Yuan2020OnePO}, items~\cite{Singh2008RelationalLV, Zhu2019DTCDRAF} or attributes~\cite{Tang2012CrossdomainCR}. To learn common item features from different domains, pre-trained language models~\cite{Devlin2019BERTPO, Liu2019RoBERTaAR} provide high-quality item features by encoding item texts (e.g.,~title, brand). Based on pre-trained item features, several methods~\cite{Ding2021ZeroShotRS, Hou2022TowardsUS} are proposed to learn universal item representations by applying additional layers. In this work, we have the same target as previous transfer learning for recommendation (i.e.,~alleviate data sparsity and cold-start item issues). However, instead of relying on common users, items and attributes or encoding items with pre-trained language models, we directly learn language representations for sequential recommendation and hence transfer knowledge based on the generality of natural languages.

%% file: 5_conclusion.tex
\section{Conclusion}
In this paper, we propose \our, a framework that can effectively learn language representations for sequential recommendation. To recommend the next item based on languages, we first formulate items as key-value attribute pairs instead of item IDs. Then, we propose a novel bi-directional Transformer model for sequence and item representations. The proposed structure can learn both natural languages and sequential patterns for recommendations. Furthermore, we design a learning framework including pre-training and finetuning that helps the model learn to recommend based on languages and transfer knowledge into different recommendation scenarios. Finally, extensive experiments are conducted to evaluate the effectiveness of \our under full-supervised and low-resource settings. Results show that \our largely outperforms existing methods in different settings, especially for the zero-shot and cold-start items recommendation which indicates \our can effectively transfer knowledge from training. An ablation study is conducted to show the effectiveness of our proposed components.